\documentstyle[12pt,epsf]{article}
\font\twelvebb=msbm10 scaled 1200
\font\tenbb=msbm10
\font\eightbb=msbm8

  \newfam\bbfam
  \def\bb{\fam\bbfam\twelvebb}
  \textfont\bbfam=\twelvebb
  \scriptfont\bbfam=\tenbb
  \scriptscriptfont\bbfam=\eightbb
\def\R{{\bb R}}			
\def\Z{{\bb Z}}			
\def\M{{\cal M}}		
\def\O{{\cal O}}		
\def\E{{\cal E}}		
\def\opt{{\hbox{\tiny opt}}}	
\def\mc{{\hbox{\tiny MC}}}	
\def\surface{{\hbox{\tiny surface}}} 
\def\<#1#2>{\ifx<#1		
  \setbox0=\hbox{$#2$}%
  \dimen0=\wd0%
  \setbox0=\hbox{$\left\langle\box0\right\rangle$}%
  \advance\dimen0by-\wd0%
  \divide\dimen0by2%
  \left\langle\kern\dimen0\,\left\langle{#2}%
    \right\rangle\,\kern\dimen0\right\rangle%
  \else%
  \left\langle{#1#2}\right\rangle%
  \fi}
\def\estimator[#1,#2]{E_{#2}\left[#1\right]} 
\def\variance[#1,#2]{V_{#2}\left[#1\right]} 
\def\freq{\omega}		
\def\w{\zeta}			
\def\force{{\cal F}}		
\def\dt{{\delta\tau}}		
\def\tz{\tau}			
\def\hb{{\hbox{\tiny HB}}}	
\def\aor{{\hbox{\tiny AOR}}}	
\def\hor{{\hbox{\tiny HOR}}}	
\def\acc{{\hbox{\tiny acc}}}	
\def\dt{{\delta\tau}}		
\def\tz{\tau}			
\def\acc{{\hbox{\tiny acc}}}	
\def\comp{{\hbox{\tiny comp}}}	
\def\draftversion{N}		
\def\note#1{\if\draftversion Y{\sl[#1]\/}\fi} 
\if \draftversion Y
\fi

\title{Progress in Lattice Field Theory Algorithms}

\author{A. D. Kennedy%
    \thanks{This research was supported by the Florida State University
      Supercomputer Computations Research Institute which is partially funded
by
      the U.S. Department of Energy through contract \#DE--FC05--85ER250000.
      This research has been carried out with the partial support of the
      Sardinia Regional Authorities. I would like to thank the University of
      Edinburgh and CRS4 (Cagliari) for their hospitality while this review was
      being written.}\\
  Supercomputer Computations Research Institute\\
  Florida State University\\
  Tallahassee FL~32306--4052, U.S.A.}

\date{October 30, 1992\\[1ex]
  \small SCRI preprint FSU--SCRI--92--160\\
  CRS4 preprint CRS4--PARCOMP--92--1\\
  hep-lat/9212017}

\begin{document}

\maketitle

\begin{abstract}
  I present a summary of recent algorithmic developments for lattice
  field theories. In particular I give a pedagogical introduction to
  the new Multicanonical algorithm, and discuss the relation between the
  Hybrid Overrelaxation and Hybrid Monte Carlo algorithms. I also attempt to
  clarify the r\^ole of the dynamical critical exponent $z$ and its
  connection with ``computational cost.''
\end{abstract}

\newpage

\section{INTRODUCTION}

In this review I shall concentrate on two areas in which there has been
considerable work and significant progress during the last year. The first
is the Multicanonical algorithm which, although the underlying ideas have
been known for some time, has been successfully applied to studying many
properties of first-order transitions in lattice field theories. I shall
attempt to give a somewhat pedagogical introduction to this method. The
second is the Hybrid Overrelaxation algorithm; this method has been used for
some years, but recent analysis of its performance for the Gaussian model
and of its relationship with the Hybrid Monte Carlo algorithm are
illuminating.

Unfortunately there has been little activity or progress on what I consider
the two outstanding algorithmic challenges facing lattice field theory
today: dynamical fermions and complex actions. For the former the algorithms
seem to work quite well, albeit slowly, whereas for the latter the situation
is much worse.

There has been a lot of work on multigrid and cluster methods. Both have
been shown to work well in some simple models (usually in two dimensions),
but their efficacy has not yet been demonstrated for four dimensional
non-abelian gauge theories. Since these methods were discussed in
considerable detail in previous lattice conferences I shall only survey the
progress very briefly here.

I shall also try to clarify the situation regarding what is known about
critical slowing down for the Hybrid Monte Carlo algorithm.

\section{MULTICANONICAL METHODS}

Although the application of the multicanonical algorithm to lattice field
theory is new \cite{berg91a,berg92a,marinari92a}, similar techniques have
been used for some time \cite{valleau77a,valleau90a,valleau91a}. For a
recent review of developments in this subject see \cite{berg92c}.

\subsection{Marginal Distributions}

A statistical mechanical system or a quantum field theory may be described
not only in terms of their detailed microscopic states, but also in terms of
a set of macroscopic parameters (e.g., energy, density, magnetization) which
suffice to specify all of its thermodynamic properties. We shall denote the
space of microstates by $\M$, which has ``volume'' $Z\equiv\int_\M d\phi$
with respect to the natural measure $d\phi$, and we shall write $\<\cdot>_f$
to indicate an average over $\M$ with respect to the measure $\propto
d\phi\,f(\phi)$. Similarly, we shall call the space of macrostates $\O$,
with volume $Z'\equiv\int_\O dE$ and averages over this space with respect
to the measure $\propto dE\,g(E)$ will be written as $\<<\cdot>_g$. If
$\Omega$ is some thermodynamic observable which depends only on some
macroscopic parameters $\E$ then the situation may be summarized by the
following equation
\begin{equation}
  \M \stackrel{\E}{\longrightarrow} \O \stackrel{\Omega}{\longrightarrow} \R.
\end{equation}

Microscopic and macroscopic averages are related in a trivial
way:\footnote{We use the notation $\Omega\circ\E$ to indicate functional
composition: $(\Omega\circ\E)(\phi)=\Omega\bigl(\E(\phi)\bigr)$.}
\begin{equation}
  \<\Omega\circ\E> = {1\over Z}\int_\M d\phi\,\Omega\bigl(\E(\phi)\bigr)
\end{equation}
\begin{equation}
  \qquad = {1\over Z} \int_\M d\phi
    \int_\O dE\,\delta\bigl(E-\E(\phi)\bigr) \Omega(E)
\end{equation}
\begin{equation}
  \qquad = \int_\O dE\,\rho(E) \Omega(E) = \<<\Omega>_\rho
\end{equation}
where we have introduced the {\em density of states}
\begin{equation}
  \rho(E) \equiv {1\over Z} \int_\M d\phi\,\delta\bigl(E-\E(\phi)\bigr).
\end{equation}
The probability distribution generated by $\rho(E)$ is the {\em marginal
distribution\/} on~$\O$, meaning that it is obtained from the full
distribution over the much larger space $\M$ by averaging over all the other
variables.

\subsection{Importance Sampling}

Importance sampling is a fundamental technique to reduce the statistical
errors in a Monte Carlo computation. Suppose we wish to measure the
expectation value of some quantity $f:\M\to\R$: to this end we introduce an
estimator
\begin{equation}
  \estimator[{f\over\mu},\mu]
    = {1\over T} \sum_{t=1}^T {f(\phi_t)\over\mu(\phi_t)}
\end{equation}
where samples $\{\phi_t\}$ are chosen from the distribution $\mu(\phi_t)$.
It is easy to see that the expectation value $\<\estimator[f/\mu,\mu]>_\mu =
\<f>$ is independent of the choice of distribution $\mu$; in fact the
central limit theorem tells us the stronger result that
\begin{equation}
  \estimator[{f\over\mu},\mu] = \<f>
    + O\left(\sqrt{\variance[f/\mu,\mu]\over T}\right),
\end{equation}
where the variance of a single sample is
\begin{eqnarray}
  \variance[{f\over\mu},\mu]
    &=& \<{\left({f\over\mu} - \<{f\over\mu}>_\mu\right)^2}>_\mu \\
    &=& \<{f^2\over\mu}> - \<f>^2.
\end{eqnarray}
Observe that we can reduce the error not only by increasing $T$ but also by
choosing $\mu$ so as to make $\variance[f/\mu,\mu]$ smaller.

We may find the optimal importance sampling by a simple variational calculation
\begin{eqnarray*}
  {\delta\over\delta\mu}
    \biggl\{ \variance[{f\over\mu},\mu]
      &+& \lambda\<\mu> \biggr\}_{\mu=\mu_\opt} = \\
  &=& - \<{f^2\over\mu_\opt^2}> + \lambda = 0.
\end{eqnarray*}
The solution of this equation gives the optimal probability distribution
\begin{equation}
  \mu_\opt={|f|\over\<|f|>};
\end{equation}
this gives a variance per sample of
\begin{equation}
  \variance[{f\over\mu_\opt},\mu_\opt] = \<|f|>^2 - \<f>^2,
\end{equation}
which vanishes iff $f$ has the same sign~$\forall\phi$. While we cannot
often achieve this ideal situation, generating a probability distribution
which approximates the operator being measured can greatly reduce the amount
of computer time necessary to get a reliable measurement of the operator's
expectation value.

\subsection{Canonical Distribution}

In field theory as well as for statistical systems we are most often
interested finding the expectation values of operators with respect to
the canonical distribution. For statistical systems this distribution
is obtained by maximizing the entropy $S\equiv-\<\log P>_P$ with
respect to the probability distribution~$P$ subject to the constraints
that the ensemble average is at a given point in $\O$: $\<1>_P=1$,
$\<\E>_P=E\in\O$.
\begin{equation}
  {\delta\over\delta P}
    \bigl\{ S + \lambda\<1>_P - \beta\<\E>_P \bigr\}_{P=P_c} = 0
\end{equation}
whose solution is
\begin{equation}
  P_c(\phi) = {e^{-\beta\E(\phi)}\over Z(\beta)},
  \label{canonical}
\end{equation}
with the {\em partition function}~$Z$ and {\em free energy}~$F$ given~by
\begin{equation}
  Z(\beta) = \int_\M d\phi\,e^{-\beta\E(\phi)} = e^{-\beta F(\beta)},
\end{equation}
and
\begin{equation}
  E = -{\partial\over\partial\beta} \log Z(\beta).
\end{equation}

In terms of macroscopic averages the canonical distribution of
Eq.~(\ref{canonical}) is
\begin{equation}
  \<\Omega\circ\E>_{e^{-\beta\E}} = \<<\Omega>_{\rho(E)e^{-\beta E}}.
\end{equation}
Taking configurations from the canonical distribution $P_c$ gives good
importance sampling if $\Omega(E)\approx1$ for all $E$ in the subspace
of $\O$ of interest.

It is important to note that we do not know $Z(\beta)$ {\em a priori}, so
using any other importance sampling requires the computation of ratios of
estimators. A ratio of unbiased estimators is not an unbiased estimator for
the ratio, so care must be taken to avoid systematic errors; as we shall
see, this ought to be done for multicanonical computations.

\subsection{Multicanonical Distribution}

\begin{figure}
  \begin{center}
    \leavevmode
    \epsfbox[200 90 770 500]{bumpy-operator.ps}
  \end{center}
  \caption{A situation in which the multicanonical method is useful. The
    solid line is the canonical distribution and the dashed line is the
    operator of interest.}
  \label{bumpy-operator}
\end{figure}

If $\Omega(E)$ has exponentially large peaks then important contributions
may come from regions where $e^{-\beta E}\rho(E)$ is small. This situation
is illustrated schematically in figure~\ref{bumpy-operator}. A compromise
to get good importance sampling simultaneously for $\Omega e^{-\beta E}$ and
$e^{-\beta E}$ is to generate configurations with a uniform density on
macrostate space~$\O$. This is the {\em multicanonical\/} distribution
\begin{equation}
  P_\mc(\phi) \propto {1\over\rho\bigl(\E(\phi)\bigr)}.
\end{equation}
It leads to $\<\Omega\circ\E>_{1\over
\rho\circ\E} = \<<\Omega>$ and
\begin{eqnarray}
  \<\Omega\circ\E>_{e^{-\beta\E}}
    &=& {\<\rho\circ\E\;e^{-\beta\E}\;\Omega\circ\E>_{1\over\rho\circ\E}
      \over \<\rho\circ\E\;\Omega\circ\E>_{1\over\rho\circ\E}} \\
    &=& {\<<\rho e^{-\beta E} \Omega> \over \<<\rho e^{-\beta E}>}.
\end{eqnarray}

\subsection{Algorithms to Generate the \protect\goodbreak Multicanonical
  Distribution}

If $\tilde\rho(E)\approx\rho(E)$ then we can readily generate an approximate
multicanonical distribution
\begin{equation}
  \tilde P_\mc(\phi) \propto{1\over \tilde\rho\bigl(\E(\phi)\bigr)}
\end{equation}
for which
\begin{eqnarray}
  \<\Omega\circ\E>_{e^{-\beta\E}}
    &=& {\<\tilde\rho\circ\E\;e^{-\beta\E}\;\
      \Omega\circ\E>_{1\over\tilde\rho\circ\E}
      \over \<\tilde\rho\circ\E\;\Omega\circ\E>_{1\over\tilde\rho\circ\E}} \\
    &=& {\<<\tilde\rho e^{-\beta E} \Omega>_{\rho/\tilde\rho}
      \over \<<\tilde\rho e^{-\beta E}>_{\rho/\tilde\rho}}.
\end{eqnarray}
This is cheap if all the macroscopic parameters $\E$ are local
operators on $\phi\in\M$. We can use a Metropolis algorithm with
acceptance probability
\begin{eqnarray}
  P(\phi\to\phi')
    &=& \min\left(1,{\tilde P_\mc(\phi')\over\tilde P_\mc(\phi)}\right) \\
    &=& \min\left(1,{\tilde\rho\bigl(\E(\phi)\bigr)
        \over \tilde\rho\bigl(\E(\phi')\bigr)}\right) \\
    &=& \min\left(1,e^{-\Delta\log\tilde\rho\circ\E}\right).
\end{eqnarray}
Presumably it would be better to use Hybrid Monte Carlo or Hybrid
Overrelaxation algorithms if they have $z\approx1$.

For low-dimensional $\O$ we may represent the approximate spectral density
$\log\tilde\rho(E)$ by binning. Berg and Neuhaus used a piecewise linear
interpolation (canonical within each bin)
\begin{equation}
  \log\tilde\rho(E) = \sum_{j=1}^N \chi_j(E) (\beta_j E + \alpha_j)
\end{equation}
[$\chi_j(E)$ is the characteristic function of bin~$j$: it has the value $1$
if $E$ is in the bin and $0$ otherwise], whereas Marinari and Parisi use a
stochastic superposition of equiprobable linear interpolations (each term
canonical)
\begin{equation}
  \tilde\rho(E) = \sum_{j=1}^N P_j (\beta_j E + \alpha_j)
\end{equation}
[$P_j$ is the probability of choosing term~$j$ from the ``sum'']. The
former is illustrated in figure~\ref{interpolation}.

\begin{figure}
  \begin{center}
    \leavevmode
    \epsfbox[200 90 770 500]{interpolation.ps}
  \end{center}
  \caption{A piecewise linear approximation to $\log\rho$.}
  \label{interpolation}
\end{figure}

\subsection{Applications}

There have been many calculations making use of the multicanonical method
during the last year, including \cite{berg92f,berg92g,berg92i,berg92b,%
berg92d,berg92h,billoire92a,grossmann92a,grossmann92b}. We shall just
consider two very simple examples in which it is effective:
\begin{itemize}
\item The surface energy at a first order phase transition, and
\item The autocorrelation time (tunnelling time) at a first order phase
  transition.
\end{itemize}

The surface energy may be defined by
\begin{equation}
  {P_{\scriptstyle\max}\over P_{\scriptstyle\min}}
    = 2 F_\surface L^{D-1} \qquad (L\to\infty),
\end{equation}
where $P_{\min}$ and $P_{\max}$ are illustrated in figure~\ref{surface-energy}.
\begin{figure}
  \begin{center}
    \leavevmode
    \epsfbox[200 90 770 500]{surface-energy.ps}
  \end{center}
  \caption{Schematic surface energy calculation.}
  \label{surface-energy}
\end{figure}
In order to relate this to our formal analysis the appropriate operators may
be expressed as
\begin{equation}
  P_{\stackrel{\max}{\min}} = \lim_{\gamma\to\pm\infty}
    {1\over\gamma} \log \<<e^{\gamma P}>;
\end{equation}
these operators are shown (for some large but finite value of $\gamma$) with
dashed lines in figure~\ref{surface-energy}. The lack of overlap between
the operator giving $P_{\min}$ and the canonical distribution is obviously
an example of the situation shown in figure~\ref{bumpy-operator}. For this
reason it is clear why this operator will be sampled much better by the
multicanonical algorithm, which will generate configurations of energy $E$
uniformly distributed over the range of interest.

\begin{figure}
  \begin{center}
    \leavevmode
    \epsfbox[200 90 770 500]{binning-error.ps}
  \end{center}
  \caption{Another situation in which the multicanonical method is useful;
    the canonical distribution at a first-order phase transition.}
  \label{binning-error}
\end{figure}
The autocorrelation time (tunnelling time) for local Monte Carlo algorithms
at a first order phase transition is notoriously long. The problem here is
that any local algorithm requires the system to pass through the minimum
separating the two phases (figure~\ref{binning-error}), so the tunnelling
time is approximately proportional to $P_{\min}$. Since the logarithm of the
density of states is an extensive quantity
\begin{equation}
  \log\rho(E) \propto L^D
\end{equation}
where $L^D$ is the lattice volume, it follows that
\begin{equation}
  \tau_A \propto P_{\min} \propto e^{-\Delta L^D},
  \label{multicanonical:z}
\end{equation}
where $\Delta$ is the error in the approximation for $\log\rho(E)$ at
any fixed volume. For a true multicanonical distribution $\Delta=0$, and for
an approximate one it is at least much smaller than the canonical barrier,
as shown in figure~\ref{binning-error}.

\subsection{Open Questions}

While the multicanonical method works very well in practice, there are still
several interesting algorithmic questions about it which ought to be
addressed.

\begin{itemize}
\item For a fixed $\tilde\rho(E)$ the autocorrelation time $\tau_A$ is
  still exponential in $L^D$, but with a much smaller exponent. This follows
  simply from equation~(\ref{multicanonical:z}).
\item We can measure $\rho(E)$ during the course of a multicanonical
  computation by counting the number of configurations landing in each
  bin, and it can be used to improve the approximation $\tilde\rho(E)$.
\item Just as for simulated annealing, however, no one has given an
  {\em algorithm} for evolving $\tilde\rho(E)$. In order to do this one
  would have to address the following issues:
  \begin{itemize}
  \item How do we distinguish a statistically significant difference between
    $\rho$ and $\tilde\rho$ from fluctuations?
  \item How do we ensure stability, and that $\tilde\rho\to\rho$?
  \item The number of bins must presumably grow as $L^D$ to avoid an
    exponential autocorrelation time, yet there must be sufficient
    statistics in each bin to give a significant estimate for $\tilde\rho$.
  \item Is there an algorithm giving a power-law volume dependence for
    $\tau_A$?
  \item Do we care? In practical simulations we might not mind having an
    exponential algorithm provided that the exponent was small enough. We
    might even prefer it to a polynomial algorithm with a large coefficient.
  \end{itemize}
\end{itemize}

\section{HYBRID OVERRELAXATION}

The name ``Hybrid Overrelaxation'' appears to have been coined by Wolff
\cite{wolff92a} who recently analyzed the dynamical critical behaviour of
this algorithm in the Gaussian model. The algorithm, however, has been
used in a variety of models over the past few years \cite{creutz87b,%
brown87a,gupta88d,gupta88e,decker89a,baillie91a,baillie91b,meyer91a,%
wolff92b,booth91a}. It belongs to a class of overrelaxed algorithms
introduced by Adler \cite{adler81a} with dynamical critical exponent
$z\approx1$ for the Gaussian model \cite{adler88b,neuberger87a,neuberger92a}.

\subsection{Adler Overrelaxation}

Consider the Gaussian model defined by the free field action
\begin{equation}
  S(\phi) = {1\over2} \sum_x\left\{
    \sum_{\mu=1}^D \bigl(\partial_\mu\phi(x)\bigr)^2 + m^2\phi(x)^2 \right\}.
\end{equation}
A single-site Adler overrelaxation (AOR) update \cite{adler81a} with
parameter $\w$ replaces
$\phi(x)$~by
\begin{equation}
  \phi'(x) = (1-\w)\phi(x) + {\w\force\over\freq^2}
    + {\sqrt{\w(2-\w)}\over\freq}\,\eta,
  \label{aor}
\end{equation}
where $\freq^2 \equiv 2D+m^2$ is the square of the highest frequency in
the spectrum, the ``force'' on $\phi(x)$ due to its neighbours is $\force
\equiv \sum_{|x-y|=1} \phi(y)$, and $\eta$ is a Gaussian-distributed
random number.

\subsection{$z=1$ for Gaussian Model}
\label{horz1}

The lattice may be updated using a checkerboard scheme, alternating the
update of all even and all odd sites. This is just a consequence of the
locality of the action. This is also the basis of Wolff's \cite{wolff92a}
analysis of the dynamical critical behaviour of overrelaxation in the
Gaussian model: the fields on the even and odd sublattices can Fourier
transformed, and equation~(\ref{aor}) becomes block diagonal in momentum
space. Determining the autocorrelations of the method thus reduce to
studying the eigenvalues of a $2\times2$ matrix.

Let us now consider two special cases of the Adler overrelaxation algorithm,
corresponding to different values of $\w$. For $\w=1$ this is the heatbath
(HB) algorithm, because $\phi'(x)$ does not depend upon~$\phi(x)$. The
exponential autocorrelation time is
\begin{equation}
  \tau_\hb = {D\over m^2+p^2} \qquad (p^2\to0),
\end{equation}
which corresponds to $z=2$. If we adjust $\w$ so as to minimize the
autocorrelation time we find that
\begin{equation}
  \w = 2 - {2m\over\sqrt D} + O(m^2), \quad\tau_\aor \approx {\sqrt D\over2m};
\end{equation}
which gives $z=1$.

\subsection{Hybrid Overrelaxation}

For $\w=2$ the new field value, $\phi'(x)$, does not depend on the noise
$\eta$, so the update is not ergodic. The Hybrid Overrelaxation (HOR)
algorithm cures this by alternating $N$ AOR steps with a single HB step.
Minimizing the autocorrelation time requires increasing $N\propto1/m$, which
leads to $\tau_\hor=1$.

Just as pointed out by Weingarten and Mackenzie
\cite{weingarten88a,mackenzie89b} for Hybrid Monte Carlo (HMC), $N$ should
be varied randomly around this value to avoid accidental non-ergodicity;
although this is probably only a problem for the Gaussian model and not for
interacting field theories.

\subsection{Overrelaxation and Hybrid \protect\goodbreak Monte Carlo}

I now want to show that the HOR algorithm --- and the AOR algorithm too ---
are closely related to the HMC algorithm. To this end lets us consider the
HMC algorithm applied to the Gaussian model: we introduce the Hamiltonian
\begin{equation}
  H(\pi,\phi) = {1\over2} \pi^2 + S(\phi)
\end{equation}
on ``fictitious'' phase space. The corresponding equation of motion is
\begin{equation}
  \ddot\phi_x = -\freq^2 \phi_x + \force,
\end{equation}
whose solution in terms of the Gaussian distibuted random initial momentum
$\pi_x$ and the initial field value $\phi_x$~is
\begin{equation}
  \phi_x(t) = \phi_x \cos\freq t
    + {1 - \cos\freq t\over\freq^2} \force
    + {\pi_x\over\freq} \sin\freq t,
\end{equation}
which is exactly the AOR update considered before if we identify $\w\equiv
1-\cos\freq t$. If we use this exact solution of the equations of motion to
generate candidate configurations for HMC then the acceptance rate will be
unity. For interacting field theories HMC provides an exact algorithm for
any value of the overrelaxation parameter~$\w$ simply by dropping the
interaction terms in the action and solving the equations of motion exactly
for the resulting Gaussian model.

Let us summarize the differences between this variant of the HMC algorithm
and the usual one. For ``conventional'' HMC:
\begin{itemize}
\item All sites are updated at once;
\item Each trajectory is of length $\tz\propto1/m$ and is a sequence of many
steps of length $\dt$;
\item There is a {\em global\/} Metropolis accept/reject step.
\end{itemize}
For ``local'' HMC \cite{rossi92b,pendleton89a,pendleton89b}:
\begin{itemize}
\item Even and odd updates are alternated;
\item Each trajectory is of length $\tz\approx\pi/\freq=O(1)$;
\item There is a {\em local\/} Metropolis accept/reject step.
\end{itemize}

\subsection{Leapfrog vs. Free Field Guidance}

It is important to realize that there are two separate issues; one is
whether a global accept/reject step is used, the other is whether an exact
solution to the free-field equations of motion or an approximate (leapfrog)
solution to the true equations of motion is used.
\begin{itemize}
\item Local Metropolis acceptance/rejection:
  \begin{itemize}
  \item Useful only for local (bosonic) theories;
  \item Acceptance rate does not depend on the lattice volume.
  \end{itemize}
\item Free-field instead of leapfrog guidance:
  \begin{itemize}
  \item Leapfrog has $\dt$ errors so $P_\acc<1$ even for free field theory;
  \item Free field guidance has errors of order $\lambda$ for interacting
    theories, whereas leapfrog has errors of order $\lambda\dt^2$, where
    $\lambda$ is the coupling constant of the interaction part of the action.
  \end{itemize}
\end{itemize}

\section{DYNAMICAL CRITICAL \protect\goodbreak EXPONENTS}

Let us begin this topic by recalling the definition of the {\em dynamical
critical exponent\/}~$z$. It relates a ``dynamical'' property of the
algorithm generating field configurations, the autocorrelation time $\tau_A$
(either the exponentional autocorrelation time or an integrated
autocorrelation time will do), and a ``static'' property of the underlying
field theory, the correlation length $\xi$:
\begin{equation}
  \tau_A\propto\xi^z\quad(\xi\to\infty).
  \label{csd}
\end{equation}
One of the main ``selling points'' of algorithms is that they reduce
($z\approx1$) or even eliminate ($z\approx0$) critical slowing down from the
value ($z\approx2$) characteristic of ``random walk'' methods. In general
terms it is possible to reduce $z$ from $2$ to~$1$ by using just the right
amount of randomness in the algorithm (see sections~\ref{horz1}
and~\ref{hmcz1}), whereas further reduction of $z$ seems \cite{neuberger92a}
to require an algorithm with more specific ``knowledge'' of the dynamics of
the theory (section~\ref{mgclz1}).

Although the theoretical analysis of critical slowing down is usually
limited to the Gaussian model, this is relevant because an algorithm is
often better just because it is not quite so dumb in dealing with free
fields. How close real theories are to free fields is an empirical question:
for some interacting field theories the values of $z$ have been determined
empirically \cite{edwards92a,gupta92a,gupta92d,vaccarino91a,sloan92a}.

\subsection{Do we care too much about $z$?}

The characterization of algorithms by their dynamical critical exponent has
perhaps been somewhat over-emphasized. Care is required not only because our
computations are not always peformed at parameters where the asymptotic
behaviour of the algorithms has set in, but also because the cost of a
computation is not just given by the autocorrelation time.

We want to study the continuum limit (critical behaviour) of some lattice
model in a large enough box (thermodynamic limit). In order that the
systematic errors are under control we need to match both the short and long
distance behaviour of the lattice regularization to some analytic form:
\begin{itemize}
\item For asymptotically free theories we hope we can match the
  $\xi/a\to\infty$ scaling to perturbation theory.
\item Non-perturbative effects fall off exponentially fast in UV asymptopia.
\item For non asymptotically free theories it is unclear how we can
  verify that $a$ is small enough.
\item We hope to match results using finite size scaling for $L/\xi\to\infty$.
\end{itemize}
We only need to carry out numerical computations for an essentially finite
range of lattice spacings. Good asymptotic dynamical scaling behaviour of an
algorithm is useful only if
\begin{itemize}
\item scaling sets in for the lattice sizes we are interested in,
\item the coefficient in equation~(\ref{csd}) is small.
\end{itemize}

\subsection{Critical Slowing Down and Finite \protect\goodbreak Size Scaling}
\label{hmcz1}

Our definition of $z$ (equation~(\ref{csd})) assumes $L\gg\xi$, which is
usually the case for lattice gauge theory computations. If we want to study
finite size scaling computations are sometimes done in the regime $L\ll\xi$,
and the dynamical critical exponent is defined by $\tau_A\propto
L^z$~\cite{mehlig92a}. It is not clear that the definitions are equivalent
if some parameters have to be tuned to values depending upon~$\xi$, e.g.,
$\tz\propto\xi$ in HMC. For HMC in the Gaussian model with $L\gg\xi$, $z=2$
if $\tz=$constant, and $z=1$ if $\tz\propto\xi$~\cite{kennedy91b,kennedy91a}.
If $\xi\gg L$ then $z=2$ for $\tz=$constant, but I know of no analysis for
any other cases.

\subsection{Computational Cost}

For the global HMC algorithm $z$ is not the whole story; the computational
cost explicitly depends on the lattice volume even at fixed~$\xi$
\cite{creutz88a,gupta88b}
\begin{equation}
  T_\comp \propto V^{5/4} = L^{5D/4}
\end{equation}
because the integration step size $\delta\tau$ has to be reduced so as to
keep the Metropolis acceptance rate constant. It must be remembered that
$T_\comp\propto V$ is true even for local algorithms (one has to look at
every site!).  For higher-order leapfrog schemes this cost can be reduced to
$T_\comp\propto V^{1+\epsilon}$ (an interesting new way of understanding
these higher-order algorithms is presented in \cite{sexton92a}).

Before leaving this subject I would like to mention a couple of related
topics. First, some interesting results on the scaling behaviour of HMC in
the presence of dynamical fermions are presented in~\cite{gupta90a,gupta92c}.
Second, an interesting variant of HMC which adds a small amount of
randomness to the momenta at each integration step (instead of a complete
momentum refreshment after an entire trajectory) was found by
Horowitz~\cite{horowitz90a}. Unfortunately, the deterministic part of the
momenta has to have its sign flipped after every step in order to satisfy
the detailed balance condition, and this means the modified algorithm does
not perform any better (although a detailed proof has not yet been
presented).

\section{MULTIGRID AND \protect\goodbreak CLUSTER METHODS}
\label{mgclz1}

I mentioned previously that algorithms which have $z<1$ appear to require
some knowledge of the detailed dynamics of the model being simulated. For
some spin models this is achieved in a non-trivial manner by cluster
algorithms, but for lattice gauge theories like QCD all attempts in this
direction are motivated by free field theory dynamics. This is not a bad
idea, because such theories are asymptotically free. There has not been much
activity in the area of Fourier acceleration, but there has been a great
deal of work on Multigrid methods. Since such methods were reviewed in
detail last year I shall merely survey work in this area extremely briefly.

Grabenstein and Pinn have presented a formalism for computing the acceptance
rate for Multigrid Monte Carlo~\cite{grabenstein92a,grabenstein92b}. There
have been numerous applications and tests of multigrid methods, including
those of Hasenbusch and Meyer~\cite{hasenbusch92a,hasenbusch92b,meyer91a}
(strictly speaking these are really ``unigrid'' algorithms), Laursen and
Vink~\cite{laursen92a}, and Edwards {\em et al.}~\cite{edwards92b}.

Multigrid methods are also under active study as a means of inverting the
lattice Dirac operator more efficiently
\cite{kalkreuter92a,hulsebos92a,lauwers90a,lauwers92a,lauwers92b}. Another
method used for this purpose has been proposed by
Vyas~\cite{vyas91a,vyas92a}. Whether these techniques can significantly
outperform the conjugate gradient algorithm is not yet settled.

Multigrid methods have also been used to expedite gauge
fixing~\cite{hulsebos92b} (see also the work of van der Sijs~\cite{sijs92a}).

Cluster algorithms have also been under active study. An interesting way of
understanding them has been suggested by Wolff~\cite{wolff91a}. Their area
of applicability is still mainly for spin models of various
kinds~\cite{evertz92b,evertz92c} where they are extremely successful and are
often the clear method of choice. Nevertheless, their use is limited to
models whose spin manifold possesses an involutive isometry whose fixed
point manifold has codimension one (i.e., a discrete quotient of a product
of spheres), as was shown by Sokal and collaborators~\cite{sokal92a}.
B.~Bunk \cite{bunk91a} describes a cluster algorithm which is applicable to
$\Z(2)$ gauge theory.

The parallelism inherent in FFT, multigrid, and cluster algorithms is more
complex than the simple grid-like structure needed for the algorithms
discussed in previous sections (they require only nearest-neighbour
communication with infrequent global summations). The implementation of
cluster algorithms on parallel and vector computers has been discussed in
several papers~\cite{flanigan92a,rossi92a,evertz92a}. In the future it will
be interesting to characterize these algorithms in terms of the type of
communications network required to implement them efficiently on a parallel
computer (e.g., combining networks, infinite dimensional grids, etc.).

\section{OTHER ALGORITHMS}

There are, of course, numerous interesting results which I have not had time
or space to discuss in detail. I do wish to mention at least a few of them
here, so that the interested reader will be able to refer to the original
literature.

An interesting algorithm called ``microcanonical cluster Monte Carlo'' was
introduced by Creutz \cite{creutz92a}. This algorithm allows a discrete spin
variable to interact with a heat bath (provided by a family of ``demons''),
and for the demon energies to be refreshed occasionally. In some ways this
method is a discrete analogue of the HMC method.

Sloan, Kusnezov, and Bulgac \cite{sloan92a} introduce a ``chaotic molecular
dynamics'' algorithm, which couples chaotic degrees of freedom to a system
to ensure ergodicity (instead of, or perhaps as well as, refreshing the
fictitious momenta of the Hybrid algorithm). Neal \cite{neal92a} suggests a
new procedure which may serve to increase the acceptance rate of the HMC
algorithm. Finally, Bhanot and Adler \cite{bhanot91a} describe a parallel
algorithm for updating spin models.



\end{document}